\documentclass[10pt]{article}
\usepackage{latexsym}
\usepackage{amssymb}
\usepackage{amsmath}
\usepackage{amscd}
\usepackage{amsthm}
\usepackage[left=2cm,top=2.5cm,right=2.5cm,bottom=1.5cm]{geometry}
\usepackage[pdftex]{graphicx}
\usepackage{hyperref}
\begin{document}
\begin{center}
\large{\bf{Viscous Dark Energy and Phantom Field in An Anisotropic Universe}} \\
\vspace{10mm}
\normalsize{Hassan Amirhashchi$^1$ and Anirudh Pradhan$^2$}\\
\vspace{5mm} \normalsize{$^1$Department of Physics, Mahshahr Branch, Islamic Azad University, Mahshahr, Iran \\
$^1$E-mail:h.amirhashchi@mhriau.ac.ir; hashchi@yahoo.com}  \\
\vspace{5mm} \normalsize{$^{2}$Department of Mathematics, Hindu
P. G. College, Zamania-232 331, Ghazipur, U. P., India \\
$^2$E-mail: pradhan@iucaa.ernet.in} \\
\end{center}
\vspace{10mm}
\begin{abstract}
In this paper we have investigated the general form of viscous and
non-viscous dark energy equation of state (EoS) parameter in the
scope of anisotropic Bianchi type I space-time. We show that the
presence of bulk viscosity causes transition of $\omega^{de}$ from
quintessence to phantom but the phantom state is an unstable state
(as expected) and EoS of DE tends to $-1$ at late time. Then we
show this phantomic description of the viscous dark energy and
reconstruct the potential of the phantom scalar field. It is found
that bulk viscosity pushes the universe to a darker region. We have
also shown that at late time $q\sim-\Omega^{de}$.
\end{abstract}
\section{Introduction}
There are observational evidences to show that our Universe is undergoing a late-time
accelerating expansion and we live in a privileged spatially flat Universe (Perlmutter
et al. 1998; Riess et al. 1998; Garnavich et al. 1998; Schmidt et al. 1998; Tonry et al. 2003;
Clocchiatti et al. 2006; de Bernardis et al. 2000; Hanany et al. 2000; Spergel et al. 2003;
Tegmark et al. 2004; Seljak et al. 2005; Adelman-MacCarthy et al. 2006; Bennett et al. 2003;
Allen et al. 2004). These observations indicate that a mysterious type of energy
called `` dark energy" which is contributing $73\%$ of the total energy of
the universe, and approximately $4\%$ baryonic matter and $23\%$
dark matter. However, the observational data are far from being
complete (for a recent review, see Perivolaropoulos 2006;
Jassal et al. 2005). In fact when the dark energy equation
of state (EoS) parameter $\omega^{(de)} = p^{(de)}/\rho^{(de)}$ is
less than $-\frac{1}{3}$, the universe exhibits accelerating
expansion. The equation of state of dark energy $\omega^{(de)}$
could be equal to $-1$ (standard $\Lambda$CDM cosmology), a little
bit upper than $-1$ (the quintessence dark energy) or less than
$-1$ (phantom dark energy) while the possibility $\omega \ll -1$
is ruled out by current cosmological data (Riess et al. 2004; Astier et al. 2006;
Eisentein et al. 2005; MacTavish et al. 2006; Komatsu et al. 2009).
There are two main candidates for dark energy (1) cosmological
constant (or vacuum energy) and (2) scalar fields. Although, a
cosmological constant can explain the current acceleration in a
natural way, but would suffer from some theoretical problems
such as fine-tuning problem and coincidence problem. Another
possible form of dark energy is provided by the dynamically
changing DE (Scalar-field dark energy models) including
quintessence, K-essence, tachyon, phantom, ghost condensate and
quintom, etc. Among these scalar fields, quintessence and phantom
are of more scientific interest. Models of dark energy with
evolving $\omega^{(de)}$ between $-\frac{1}{3}$  and $-1$ are
refereed to quintessence. But as a candidate for dark energy,
quintessence field with $\omega^{(de)}>-1$ is not consistent with
the recent observations which indicate that $\omega^{(de)}<-1$ (at
$z\sim 0.2$) is allowed at $68\%$ confidence level. Models with
$\omega^{de}<-1$ introduce a scalar field $\phi$ that is minimally
coupled to gravity with a negative kinetic energy and are known as
``Phantom fields" (Caldwell 2002). Unfortunately, phantom fields are
generally plagued by ultraviolet quantum instabilities (Carroll et al. 2003).\\

The negative pressure of the dark energy may be the cause of the
acceleration of the present Universe. However, the nature of the
dark energy still remains a mystery. No more than eight
years ago, some physicists (McInnes 2002; Barrow 2004) found that,
if we assumed the cosmic fluid to be ideal only, i.e. non-viscous, it
must bring out the occurrence of a singularity of the universe in
the far future. There are two methods to modify or soften the
singularity. The first is the effect of quantum corrections due to
the conformal anomaly (Brevik and Odintsov 1999; Nojiri and Odintsov
2003, 2004). The other is to consider the bulk viscosity of the cosmic
fluid (Brevik and Hallanger 2004). The viscosity theory of relativistic
fluids was first suggested by Eckart (1940), Landau and Lifshitz (1987).
The introduction of viscosity into cosmology has been investigated from
different view points (Gr$\o$n 1990; Padmanabhan and Chitre 1987; Barrow
1986; Zimdahl 1996; Maartens 1996). The astrophysical observations also
indicate some evidences that cosmic media is not a perfect fluid (Jaff et al.
2005), and the viscosity effect could be concerned in the evolution of the
universe (Brevik and Gorbunova 2005; Brevik et al. 2005; Cataldo et al. 2005).
It was also argued in (Zimdahl et al. 2001; Balakin et al. 2003), that a
viscous pressure can play the role of an agent that drives the present
acceleration of the Universe. The possibility of a viscosity dominated late
epoch of the Universe with accelerated expansion has already been mentioned by
Padmanabhan and Chitre (1987).\\

Brevik and Gorbunova (2005), Oliver et al (2011), Chen et al (2011), Jamil and
Farooq (2010), Sheykhi and Setare (2010) and Amirhashchi (2013 a,b) have studied
viscous dark energy models in different contexts. Recently, viscous dark energy
and generalized second law of thermodynamics has been studied by Setare and
Sheykhi (2010). Nojiri and Odintsov (2005) studied the effect of modification
of general equation of state (EoS) of dark energy ideal fluid by the insertion
of inhomogeneous, Hubble parameter dependent term in the late-time universe.
They also described several explicit examples of such term which is motivated
by time-dependent bulk viscosity or deviations from general relativity. The
inhomogeneous term in EoS helps to realize FRW cosmologies admitting the crossing
of phantom barrier in a more natural way. Brevik et al. (2010) have also derived a
Cardy-Verlinde (CV) formula in FRW universe with inhomogeneous generalized fluid
(including viscous fluid). They have also investigated the universality of the
dynamical entropy bound near a future singularity as well as near the Big Bang
singularity. In the present paper, first we show that the equation of state of
dark energy can cross the phantom divided line, $\omega=-1$, by introducing bulk
viscosity into the cosmic fluid but this state ($\omega^{(de)}<-1$) is a temporary
phase since the viscosity is a decreasing function of time then we suggest a
correspondence between the viscous dark energy scenario and the phantom dark energy
models in an anisotropic space-time. We show this phantomic description of the viscous
in the scope of Bianchi type I universe, and reconstruct the potential of the
phantom scalar field.
\section{The Metric and Field  Equations}
Although the FLRW models are very successful in explaining the
major features of the observed universe but the real universe is
not FLRW because of all the structure it contains, and because of
the non-linearity of Einstein's field equations the other exact
solutions we attain have higher symmetry than the real
universe. Thus, in order to obtain realistic models we can compare
detailed observations aiming to obtain `almost FLRW' models
representing a universe that is FLRW-like on large scales but
allowing for generic inhomogeneities and anisotropies arising
during structure formation on a small scale. Such models are given
by the so called ``Bianchi Type Space-Times'' which are
homogeneous but anisotropic. Goliath and Ellis (1999) have
shown that some Bianchi models isotropise due to inflation.\\

For the propose of this paper in this section we consider the
Bianchi type I space-time in the orthogonal form as
\begin{equation}
\label{eq1}
ds^{2} = -dt^{2} + A^{2}(t)dx^{2}+B^{2}(t)dy^{2}+C^{2}(t)dz^{2},
\end{equation}
where $A(t), B(t)$ and $C(t)$ are functions of cosmic time only. \\

The Einstein's field equations ( in gravitational units $8\pi G = c = 1 $) read as
\begin{equation}
\label{eq2} R^{i}_{j} - \frac{1}{2} R g^{i}_{j} = T^{(m)i}_{j} +
T^{(de)i}_{j},
\end{equation}
where $T^{(m)i}_{j}$ and $T^{(de)i}_{j}$ are the energy momentum tensors of barotropic matter and dark energy,
respectively. These are given by
\[
  T^{(m)i}_{j} = \mbox{diag}[-\rho^{(m)}, p^{(m)}, p^{(m)}, p^{(m)}],
\]
\begin{equation}
\label{eq3} ~ ~ ~ ~ ~ ~ ~ ~  = \mbox{diag}[-1, \omega^{(m)}, \omega^{(m)}, \omega^{(m)}]\rho^{m},
\end{equation}
and
\[
 T^{(de)i}_{j} = \mbox{diag}[-\rho^{(de)}, p^{(de)}, p^{(de)}, p^{(de)}],
\]
\begin{equation}
\label{eq4} ~ ~ ~ ~ ~ ~ ~ ~ ~ ~ ~ ~ ~ ~ = \mbox{diag}[-1, \omega^{(de)}, \omega^{(de)},
\omega^{(de)}]\rho^{(de)},
\end{equation}
where $\rho^{(m)}$ and $p^{(m)}$ are, respectively, the energy density and pressure of the perfect fluid
component or ordinary baryonic matter while $\omega^{(m)} = p^{(m)}/\rho{(m)}$ is its EoS parameter. Similarly,
$\rho^{(de)}$ and $p^{(de)}$ are, the energy density and pressure of the DE component respectively while
$\omega^{(de)} = p^{(de)}/\rho^{(de)}$ is the corresponding EoS parameter. We assume the four velocity vector
$u^{i} = (1, 0, 0, 0)$ satisfying $u^{i}u_{j} = -1$. \\

In a co-moving coordinate system ($u^{i} = \delta^{i}_{0}$), Einstein's field equations (\ref{eq2}) with
(\ref{eq3}) and (\ref{eq4}) for B-I metric (\ref{eq1}) lead to the following system of equations:

\begin{equation}
\label{eq5} \frac{\ddot{B}}{B}+\frac{\ddot{C}}{C}+\frac{\dot{B}\dot{C}}{BC}=-\omega^{m}\rho^{m}-\omega^{de}\rho^{de},
\end{equation}
\begin{equation}
\label{eq6} \frac{\ddot{A}}{A}+\frac{\ddot{C}}{C}+\frac{\dot{A}\dot{C}}{AC}=-\omega^{m}\rho^{m}-\omega^{de}\rho^{de},
\end{equation}
\begin{equation}
\label{eq7}\frac{\ddot{A}}{A}+\frac{\ddot{B}}{B}+\frac{\dot{A}\dot{B}}{AB}=-\omega^{m}\rho^{m}-\omega^{de}\rho^{de},
\end{equation}
\begin{equation}
\label{eq8} \frac{\dot{A}\dot{B}}{AB}+\frac{\dot{A}\dot{C}}{AC}+\frac{\dot{B}\dot{C}}{BC}=\rho^{m}+\rho^{de}.
\end{equation}
If we consider $a=(ABC)^{\frac{1}{3}}$ as the average scale factor of Bianchi type I model then
the generalized mean Hubble's parameter $H$ defines as
\begin{equation}
\label{eq9} H = \frac{\dot{a}}{a} = \frac{1}{3}\left(\frac{\dot{A}}{A} + \frac{\dot{B}}{B} + \frac{\dot{C}}{C}\right).
\end{equation}

The Bianchi identity $G^{;j}_{ij} = 0$ leads to  $T^{;j}_{ij} = 0$. Therefore, the continuity equation
for dark energy and baryonic matter can be written as
\begin{equation}
\label{eq10} \dot{\rho}^{m} + 3H(1 + \omega^{m})\rho^{m} + \dot{\rho}^{de} + 3H(1 + \omega^{de})\rho^{de} = 0.
\end{equation}
\section{Dark Energy Equation of State}
In this section we obtain the general form of the equation of
state for the viscous and non viscous energy density $\rho^{de}$
in Bianchi type I space-time when there is no interaction between
dark energy density and a Cold Dark Matter( CDM) with
$\omega_{m}=0$. But before this, we drive the general solution for
the Einstein's field equations (\ref{eq5})-(\ref{eq8}).\\

Using the method introduced by Saha (2005), when Eq. (\ref{eq5}) is subtracted from Eq. (\ref{eq6}), Eq. (\ref{eq6})
from Eq. (\ref{eq7}), and Eq. (\ref{eq5}) from Eq. (\ref{eq7}) we
obtain
\begin{equation}
\label{eq11} \frac{\ddot{A}}{A} - \frac{\ddot{B}}{B} +
\frac{\dot{C}}{C}\left(\frac{\dot{A}}{A} -
\frac{\dot{B}}{B}\right) = 0,
\end{equation}

\begin{equation}
\label{eq12} \frac{\ddot{B}}{B} - \frac{\ddot{C}}{C} +
\frac{\dot{A}}{A}\left(\frac{\dot{B}}{B} -
\frac{\dot{C}}{C}\right) = 0,
\end{equation}
and
\begin{equation}
\label{eq13} \frac{\ddot{A}}{A} - \frac{\ddot{C}}{C} +
\frac{\dot{B}}{B}\left(\frac{\dot{A}}{A} -
\frac{\dot{C}}{C}\right) = 0.
\end{equation}
First integral of Eqs. (\ref{eq11}), (\ref{eq12}) and (\ref{eq13})
leads to
\begin{equation}
\label{eq14}
\frac{\dot{A}}{A}-\frac{\dot{B}}{B}=\frac{k_{1}}{ABC},
\end{equation}
and
\begin{equation}
\label{eq15}
\frac{\dot{B}}{B}-\frac{\dot{C}}{C}=\frac{k_{2}}{ABC},
\end{equation}
\begin{equation}
\label{eq16}
\frac{\dot{A}}{A}-\frac{\dot{C}}{C}=\frac{k_{3}}{ABC},
\end{equation}
where $k_{1}$, $k_{2}$ and $k_{3}$ are constants of integration.
By taking integral from Eqs. (\ref{eq14}), (\ref{eq15}) and
(\ref{eq16}) we get
\begin{equation}
\label{eq17} \frac{A}{B}=d_{1}exp[k_{1}\int(ABC)^{-1}dt],
\end{equation}

\begin{equation}
\label{eq18} \frac{\dot{B}}{C}=d_{2}exp[k_{2}\int(ABC)^{-1}dt],
\end{equation}
and
\begin{equation}
\label{eq19} \frac{\dot{A}}{C}=d_{3}exp[k_{3}\int(ABC)^{-1}dt]
\end{equation}
where, $d_{1}, d_{2}$ and $d_{3}$ are constants of integration.\\
Now, we can find all metric potentials from Eqs. (\ref{eq17}),
(\ref{eq19}) as follow
\begin{equation}
\label{eq20} A(t)=a_{1}a~ exp(b_{1}\int a^{-3}dt),
\end{equation}
\begin{equation}
\label{eq21} B(t)=a_{2}a~ exp(b_{2}\int a^{-3}dt),
\end{equation}
and
\begin{equation}
\label{eq22} C(t)=a_{3}a~ exp(b_{3}\int a^{-3}dt).
\end{equation}
Here
\[
a_{1}=(d_{1}d_{2})^{\frac{1}{3}},~~~~~a_{2}=(d_{1}^{-1}d_{3})^{\frac{1}{3}},~~~~~a_{3}=(d_{2}d_{3})^{-\frac{1}{3}}
,
\]
\[
b_{1}=\frac{k_{1}+k_{2}}{3},~~~~~b_{2}=\frac{k_{3}-k_{1}}{3},~~~~~b_{3}=-\frac{k_{2}+k_{3}}{3},
\]
where
\[
a_{1}a_{2}a_{3}=1,~~~~~~~b_{1}+b_{2}+b_{3}=0.
\]

Therefore, one can write the general form of Bianchi type I metric
as
\begin{equation}
\label{eq23} ds^{2}=-dt^{2}+a^{2}\left[a_{1}^{2} e^{2b_{1}\int
a^{-3}dt}dx^{2}+a_{2}^{2} e^{2b_{2}\int a^{-3}dt}dy^{2}+a_{3}^{2}
e^{2b_{3}\int a^{-3}dt}dz^{2}\right].
\end{equation}
In case of non-interacting two fluid the conservation equation
(\ref{eq10}) for dark and barotropic fluids can be written
separately as
\begin{equation}
\label{eq24} \dot{\rho}^{de} + 3H(1 + \omega^{de})\rho^{de} = 0,
\end{equation}
and
\begin{equation}
\label{eq25} \dot{\rho}^{m} + 3H\rho^{m}=0.
\end{equation}
Using Eqs. (\ref{eq14}) and (\ref{eq15}) in (\ref{eq8}), we can
write the analogue of the Friedmann equation as
\begin{equation}
\label{eq26} \rho = 3H^{2} - \sigma^{2},
\end{equation}
where $\rho = \rho^{m} + \rho^{de}$ is the total energy density
and $\sigma^{2} = \frac{b_{1}b_{2} + b_{1}b_{3}
+ b_{2}b_{3}}{3a^{6}}$.  \\

Differentiating Eq. (\ref{eq26}) with respect to the cosmic time
$t$, we get
\begin{equation}
\label{eq27} \dot{\rho} = 6H\dot{H}-2\sigma\dot{\sigma}.
\end{equation}
Using Eqs. (\ref{eq24}) and (\ref{eq25}) we get
\begin{equation}
\label{eq28} \dot{\rho} = -3H(1 + \omega)\rho,
\end{equation}
where
\begin{equation}
\label{eq29} \omega = \frac{\omega^{de}\rho^{de}}{\rho} =
\frac{\omega^{de}\rho^{de}}{3H^{2}\Omega^{de}},
\end{equation}
and $\Omega^{de}=\frac{\rho^{de}}{3H^{2}}$. \\

On substituting $\dot{\rho}$ from Eq. (\ref{eq27}) into Eq.
(\ref{eq29}) we obtain
\begin{equation}
\label{eq30} \omega = -1 - 2\left(\frac{\dot{H}-Q}{\rho}\right).
\end{equation}
Using Eqs. (\ref{eq29}) and (\ref{eq30}), we can rewrite the dark
energy equation of state parameter as
\[
\omega^{de} = -\left[1 + r +
2\left(\frac{\dot{H}-Q}{3H^{2}\Omega^{de}}\right)\right]
\]
\begin{equation}
\label{eq31}
=-\left[1+r+\frac{2}{3\Omega^{de}}(\sigma^{2}-(q+1))\right],
\end{equation}

Here $q$ is the deceleration parameter (see Eq. (\ref{eq49}), $r =
\frac{\rho^{m}}{\rho^{de}}$ and $Q =
\frac{\sigma\dot{\sigma}}{3H}$. We note that since always
$\dot{\sigma} < 0$ then $Q < 0$. Also since there is no
interaction between Dark energy and CDM, $r$ is a decreasing function of time.\\

Based on the recent observations the deceleration parameter is
restricted as $-1\leq q < 0$. Therefore, from Eq. (\ref{eq31}) we
observe that the minimum value of $ \omega^{de}$ which could be
achieved for non-viscous DE is $-1$ i.e EoS of non-viscous DE cannot
cross the phantom divided line (PDL) and always varying in
quintessence region. Also from this equation we observe that at
present time i.e for $r_{0}\simeq 0.43$, $\sigma^{2}_{0}\sim 0$,
$q_{0}\simeq-0.55$, and $\Omega^{de}_{0}=0.7$,
$\omega^{de}_{0}\simeq-0.57$. But as mentioned before, according
to the current observational data the possibility of
$\omega^{de} < -1$ (crossing PDL) is allowed at $66\%$ confidence
level.  In what follows we show that by assuming a viscous DE,
$\omega^{de}$ of Eq. (\ref{eq31}) crosses PDL i.e there is
transition from quintessence to phantom region if viscosity is considered.\\

In Eckart's theory (1940) a viscous dark energy EoS parameter is
specified by
\begin{equation}
\label{eq32} {p}^{de}_{eff} = p^{de} + \Pi.
\end{equation}
Here $\Pi = -\xi(\rho^{de})u^{i}_{;i}$ is the viscous pressure and
$H = \frac{u^{i}_{;i}}{3}$ is the Hubble's parameter. On
thermodynamical grounds, in conventional physics $\xi$ has to be
positive. This is a consequence of the positive sign of the change in
entropy as an irreversible process (Landau and Lifshitz 1987). In
general, $\xi(\rho^{de})=\xi_{0}(\rho^{de})^{\tau}$, where
$\xi_{0}>0$ and $\tau$ are constant parameters. A power-law expansion for
the scale factor can be achieved for $\tau=\frac{1}{2}$ \cite{ref49}. It is
worth to mention that the Eckart's theory may suffer from causality problem
since it only consider the first-order deviation from equilibrium, however,
one can still apply it to phenomena which are quasi-stationary, i.e. slowly varying
on space and time characterized by the mean free path and the mean collision time.\\

From Eq. (\ref{eq32}) we obtain
\begin{equation}
\label{eq33} \omega^{de}_{eff} = \omega^{de} +
\frac{\Pi}{\rho^{de}}.
\end{equation}
Using Eq. (\ref{eq31}), above equation can be written as
\[
\omega^{de}_{eff} = -\left[1 + r + 2\left(\frac{\dot{H} -
Q}{3H^{2}\Omega^{de}}\right)\right] -
\xi_{0}\sqrt{\frac{3}{\Omega^{de}}}
\]
\begin{equation}
\label{eq34}=
-\left[1+r+\frac{2}{3\Omega^{de}}(\sigma^{2}-(q+1))\right]-
\xi_{0}\sqrt{\frac{3}{\Omega^{de}}},
\end{equation}
where we have assumed that $\xi(\rho^{de})=\xi_{0}\sqrt{\rho^{de}}$.
This is the general form of the viscous dark energy equation of state
in Bianchi type-I space-time. From Eq. (\ref{eq34}), we observe that
$\omega^{de}_{eff}<-1$ (cross PDL) if viscosity is considered. It is
obvious that $\omega^{de}$ tends to $-1$ as $\xi(\rho^{de})$ vanishes
at late time.\\

Eq. (\ref{eq34}) implies that one can generate phantom-like
equation of state from viscous dark energy model in Bianchi type I
universe. Thus, we assume that a phantom scalar field $\phi$ is
the origin of the dark energy. Therefore,
\begin{equation}
\label{eq35} \rho_{\phi} = -\frac{1}{2}\dot{\phi}^{2} + V(\phi),
\end{equation}
\begin{equation}
\label{eq36} p_{\phi} = -\frac{1}{2}\dot{\phi}^{2} - V(\phi).
\end{equation}
Thus, $\omega^{de}$ is given by
\begin{equation}
\label{eq37} \omega^{de}_{eff} = -\frac{V(\phi) +
\frac{1}{2}\dot{\phi}^{2}}{V(\phi) - \frac{1}{2}\dot{\phi}^{2}}.
\end{equation}
We observe that in this case $\omega^{de} < -1$. Therefore,
according to Eqs. (\ref{eq37}) and (\ref{eq34}), in the scope of
Bianchi type I universe, both non-viscous and viscous dark energy
can always be described by phantom. Eqs. (\ref{eq35}) and
(\ref{eq36}) also can be written as
\begin{equation}
\label{eq38} V(\phi) = \frac{1}{2}(1 -
\omega^{de}_{eff})\rho^{de},
\end{equation}
\begin{equation}
\label{eq39} \dot{\phi}^{2} = -(1+\omega^{de}_{eff})\rho^{de}.
\end{equation}
Using $\omega^{de}_{eff}$ from Eq. (\ref{eq34}) in to Eqs.
(\ref{eq38}) and (\ref{eq39}) we obtain
\begin{equation}
\label{eq40} V(\phi) = \frac{3H^{2}}{2}\left[(r + 2)\Omega^{de}
-\frac{1}{3} \left(1+q+\frac{Q}{H^{2}}\right) +
\xi_{0}\sqrt{3\Omega^{de}}\right].
\end{equation}
\begin{equation}
\label{eq41} \dot{\phi}^{2} = 3H^{2}\left[r\Omega^{de}
-\frac{1}{3} \left(1+q+\frac{Q}{H^{2}}\right) +
\xi_{0}\sqrt{3\Omega^{de}}\right].
\end{equation}
Now, according to Ref (Alam et al. 2004), we assume the following scalar
field equation
\begin{equation}
\label{eq42} -\ddot{\phi} - 3H\dot{\phi}^{2} + V'(\phi)=0.
\end{equation}
The solution of above equation leads to
\begin{equation}
\label{eq43} \phi = t,~~~~~~~~~ H = f(t),
\end{equation}
which implies that $f(\phi)$ must satisfy following condition
\begin{equation}
\label{eq44} 3f(\phi) = V'(\phi).
\end{equation}
We can define $\dot{\phi}^{2}$ and $V (\phi)$ in terms of
single function $f(\phi)$ also as (Nojiri and Odintsov 2006)
\begin{equation}
\label{eq45} V(\phi) = V(\phi) = \frac{3f(\phi)^{2}}{2}\left[(r +
2)\Omega^{de} + \left(\frac{f'(\phi) - Q} {3f(\phi)^{2}}\right) +
\xi_{0}\sqrt{3\Omega^{de}}\right],
\end{equation}
\begin{equation}
\label{eq46} 1 = 3f(\phi)^{2}\left[r\Omega^{de} +
\left(\frac{f'(\phi) - Q}{3f(\phi)^{2}}\right) +
\xi_{0}\sqrt{3\Omega^{de}}\right].
\end{equation}
From Eq. (\ref{eq46}) we can find $\Omega^{de}$ as
\begin{equation}
\label{eq47} \Omega^{de} = \frac{-3\xi^{2}_{0} +
\sqrt{9\xi^{4}_{0}+4r^{2}\left(\frac{1 - f'(\phi) + Q}
{3f(\phi)}\right)^{2}}}{2r^{2}}.
\end{equation}
Substituting the above $\Omega^{de}$ into Eq. (\ref{eq45}), we
obtain the scalar potential as following
\begin{equation}
\label{eq48}
V(\phi) =
\frac{3f(\phi)^{2}}{2}\left[(\frac{r+2}{2r^{2}})\left(-3\xi^{2}_{0}
+ \sqrt{9\xi^{4}_{0} + 4r^{2}\Gamma^{2}}\right) +
\frac{1}{3f(\phi)^{2}} - \Gamma+\sqrt{1.5}
\frac{\xi_{0}}{r}\sqrt{-3\xi^{2}_{0} + \sqrt{9\xi^{4}_{0} +
4r^{2}\Gamma^{2}}}\right],
\end{equation}
where $\Gamma = \frac{1 - f'(\phi) + Q}{3f(\phi)^{2}}$.\\

For completeness, we give the deceleration parameter
\begin{equation}
\label{eq49} q = -\frac{\ddot{a}}{aH^{2}} = -1 -
\frac{\dot{H}}{H^{2}},
\end{equation}
which combined with the Hubble parameter and the dimensionless
density parameters form a set of useful parameters for the
description of the astrophysical observations. From eqs.
(\ref{eq26})-(\ref{eq28}) and (\ref{eq30}), we obtain
\begin{equation}
\label{eq50} \frac{\dot{H}}{H^{2}} =
\frac{1}{3}\frac{\sigma\dot{\sigma}}{H^{2}} + 2(1 + r)(\dot{H} -
\sigma \dot{\sigma})\Omega^{de}.
\end{equation}
Using Eq. (\ref{eq50}) in Eq. (\ref{eq49}), we get
\begin{equation}
\label{eq51} q = -\left[(1+r)\Omega^{de}(1 + 2\dot{H} -
2\sigma\dot{\sigma}) + \frac{\sigma(\sigma + \dot{\sigma})}
{3H^{2}}\right].
\end{equation}
Above equation shows that at late time, $q\sim-\Omega^{de}$.
\section{Late Time Geometry of The Model}
From geometrical point of view, all FLRW based cosmological models are
homogeneous and isotropic. It is clear that such models can not describe
the evolution of our universe in it's early times where, geometrically, it
was inhomogeneous. Also, according to the recent observations, there is
tiny variations between the intensities of the microwaves coming from
different directions which means that our current universe is anisotropic.
Moreover, as far as we use the maximally symmetric FLRW metrics, one can
always ask: does the universe necessarily have the same symmetries on
very large scales outside the particle horizon or at early times? Hence,
to be more general, it is quiet reasonable to use generalized FLRW equations
by considering an anisotropic metric (Bianchi Models). \\

To show that how Bianchi models tend to isotropy, we define the generalized mean Hubble’s parameter $H$ as
\begin{equation}
\label{eq52} H= \frac{1}{3}\left(H_{1}+H_{2}+H_{3}\right),
\end{equation}
where $H_{1}=\frac{\dot{A}}{A},~H_{2}=\frac{\dot{B}}{B},~H_{3}=\frac{\dot{C}}{C}$ are the directional Hubble’s
parameters in the directions of $x$, $y$, and $z$ respectively. The mean anisotropy parameter $A_{m}$ is given by
\begin{equation}
\label{eq53} A_{m}= \frac{1}{3}\sum_{i=1}^{3}\left(\frac{\triangle H_{i}}{H}\right)^{2},
\end{equation}
where $\triangle H_{i}=H-H_{i}$. Using Eqs. (\ref{eq20})$-$(\ref{eq22}) and (\ref{eq52}) in Eq. (\ref{eq53}), we obtain
\begin{equation}
\label{eq54} A_{m}= \frac{1}{3}K \frac{a^{-6}}{1+a^{-6}},
\end{equation}
where $K=b_{1}^{2}+b_{2}^{2}+b_{3}^{2}$. Since $a=(1+z)^{-1}$, we can re-write the above equation in terms
of redshift $z$ as
\begin{equation}
\label{eq55} A_{m}= \frac{1}{3}K \frac{(1+z)^{6}}{1+(1+z)^{6}}.
\end{equation}
Equation (\ref{eq55}) obviously shows that at late time i.e $z\to -1$, $A_{m}\to 0$. Also since for
$K=0~( \mbox{i.e}~ b_{1}=b_{2}=b_{3}=0)$ our model is equivalent to the FLRW model as from
Eq. (\ref{eq55}), we obtain $A_{m}=0$.
\section{Conclusion}
Models with $\omega^{de}$ crossing $-1$ near the past have been
mildly favored by the analysis on the nature of dark energy from
recent observations (for example see Astier et al. 2006).  SNe Ia alone
favors a $\omega$ larger than $-1$ in the recent past and less
than $-1$ today, regardless of whether using the thesis of a flat
universe (Astier et al. 2006; Nojiri and Odintsov 2006) or not (Dicus and Repko 2004).
In this paper, we have studied the possibility of crossing phantom divided line
($\omega^{de}$=-1) in the scope of anisotropic Bianchi type I
space-time. The general form of the EoS parameter of viscous and
non-viscous dark energy has been investigated. It is found that
the presence of bulk viscosity causes transition of $\omega^{de}$
from quintessence to phantom. But since
$\xi(\rho^{de})=\xi_{0}\sqrt{\rho^{de}}$ and $\rho^{de}$ is a
decreasing function of time in an expanding universe we conclude
that the bulk viscosity dies out as time goes on. In another words,
the phantom state is an unstable state (as expected) and EoS of DE
tends to $-1$ at late time. It is worth to mention that equations
(\ref{eq5})$-$(\ref{eq8}) can be recast in terms of $H$, $\Sigma$
and $q$ as
\begin{equation}
\label{eq56} \bar{p} = H^{2}(2q-1) - \Sigma^{2},
\end{equation}
\begin{equation}
\label{eq57} \rho = 3H^{2} - \Sigma^{2}.
\end{equation}
Here $\Sigma^{2}$ is the shear scalar which is given by
\begin{equation}
\label{eq58} \Sigma^{2} = \frac{1}{2}\Sigma^{ij}\Sigma_{ij},
\end{equation}
where
\[
\Sigma_{ij} = u_{i;j} + \frac{1}{2}(u_{i;k}u^{k}u_{j} +
u_{j;k}u^{k}u_{i}) + \frac{1}{3}\theta(g_{ij} + u_{i}u_{j}).
\]
From equations (\ref{eq57})-(\ref{eq58}), we obtain
\begin{equation}
\label{eq59} \frac{\ddot{a}}{a} = \frac{1}{2}\xi\theta -
\frac{1}{6}(\rho^{de} + 3p^{de}) - \frac{1}{6}(\rho^{m} + 3p^{m})
- \frac{2}{3}\Sigma^{2},
\end{equation}
which is Raychaudhuri's equation for given distribution. Above
equation can be written as
\begin{equation}
\label{eq60} \frac{\ddot{a}}{a} = \frac{1}{2}\xi \theta -
\frac{1}{6}\rho^{de}(1 + 3\omega^{de}) - \frac{1}{6}\rho^{m}(1 +
3\omega^{m}) - \frac{2}{3}\Sigma^{2}.
\end{equation}
Equation (\ref{eq59}) shows that for $\rho^{de} + 3p^{de} = 0$,
acceleration is initiated by bulk viscosity only. In absence of bulk
viscosity dark energy contributes the acceleration (since $\omega^{de} < -1$,
then $1 + 3\omega^{de} < -1$, i.e the second term in the right hand side
of Eq. (\ref{eq60}) is always positive). Therefore, the presence of bulk
viscosity pushes the universe to a darker region.
\section*{Acknowledgments}
This work has been supported by a research fund from the Mahshahr Branch
of Islamic Azad University under the project entitled ``The role of scalar
fields in the study of dark energy". A. Pradhan acknowledges the financial support in part by the University Grants Commission, New Delhi, India under the grant Project F.No. 41-899/2012 (SR). The authors thank the anonymous referee
for his valuable comments.


\begin{thebibliography}{000}
\bibitem {ref1}
Adelman-McCarthy, J.K., et al.: Astrophys. J. Suppl. {\bf 162}, 38 (2006)
\bibitem {ref2}
Alam, U., Sahni, V. Starobinsky, A.A.: JCAP. {\bf 0406}, 008 (2004)
\bibitem {ref3}
Allen, S.W., et al., Mon. Not. R. Astron. Soc. {\bf 353}, 457 (2004)
\bibitem {ref4}
Amirhashchi, H.: Astrophys. Space Sci. {\bf 345}, 439 (2013a)
\bibitem {ref5}
Amirhashchi, H.: Astrophys. Space Sci. DOI: 10.1007/s10509-013-1675-z (2013b)
\bibitem {ref6}
Astier, P., et al.: Astron. Astrophys. {\bf 447}, 31 (2006)
\bibitem {ref7}
Balakin, A.B., Pav\'{o}n, D., Schwarz, D.J., Zimdahl, W.: New. J. Phys. {\bf 5}, 85 (2003)
\bibitem {ref8}
Barrow, J.D.: Class. Quantum Grav. {\bf 21}, L79 (2004)
\bibitem {ref9}
Barrow, J.D.: Phys. Lett. B. {\bf 180}, 335 (1986)
\bibitem {ref10}
Bennett, C.L., et al.: Astrophys. J. Suppl. {\bf 148}, 1 (2003)
\bibitem {ref11}
Brevik, I., Gorbunova, O.: Gen. Relat. Grav. {\bf 37}, 2039 (2005)
\bibitem {ref12}
Brevik, I., Gorbunova, O., Shaido, Y.A.: Int. J. Theor. Phys. D. {\bf 14}, 1899 (2005)
\bibitem {ref13}
Brevik, I., Hallanger.: Phys. Rev. D. {\bf 69}, 024009 (2004)
\bibitem {ref14}
Brevik, I., Nojiri, S., Odintsov, S.D., Saez-Gomez, D.: Eur. Phys. J. C. {\bf 69}, 563 (2010)
\bibitem {ref15}
Brevik, I., Odintsov, S.D.: Phys. Lett. B. {\bf 455}, 104 (1999)
\bibitem {ref16}
Caldwell, R.R.: Phys. Lett. B. {\bf 545}, 23 (2002)
\bibitem {ref17}
Carroll, S.M., Hoffman, M., Trodden, M.: Phys. Rev. D. {\bf 68}, 023509 (2003)
\bibitem {ref18}
Cataldo, M., Cruz, N., Lepe, S.: Phys. Lett. B. {\bf 619}, 5 (2005)
\bibitem {ref19}
Chen, J., Zhou, S., Wang, Y.: Chin. Phys. Lett. {\bf 28 }, 029801 (2011)
\bibitem {ref20}
Clocchiatti, A., et al.: Astrophys. J. {\bf 642}, 1 (2006)
\bibitem {ref21}
De Bernardis, P., et al., Nature. {\bf 404}, 955 (2000)
\bibitem {ref22}
Dicus, D.A., Repko, W.W.: Phys. Rev. D. {\bf 70}, 083527 (2004)
\bibitem {ref23}
Eckart, C.: Phys. Rev. {\bf 58 }, 919 (1940)
\bibitem {ref24}
Eisentein, D.J., et al.: Astrophys. J. {\bf 633}, 560 (2005)
\bibitem {ref25}
Garnavich, P.M., et al.: Astrophys. J. {\bf 509}, 74 (1998)
\bibitem {ref26}
Goliath, M., Ellis, G.F.R.: Phys. Rev. D. {\bf 60},032502 (1999)
\bibitem {ref27}
Gr$\o$n, $\O$.: Astrophys. Space Sci. {\bf 173}, 191 (1990)
\bibitem {ref28}
Hanany, S., et al.: Astrophys. J. {\bf 545}, L5 (2000)
\bibitem {ref29}
Jaffe, T.R., Banday, A.J., Eriksen, H.K., G$\grave{o}$rski, K.M., Hansen, F.K.: Astrophys. J. {\bf 629}, L1 (2005)
\bibitem {ref30}
Jamil, M., Umar Farooq, M.: Int. J. Theor. Phys. {\bf 49}, 42 (2010)
\bibitem {ref31}
Jassal, H., Bagla, J., Padmanabhan, T.: Phys. Rev. D. {\bf 72}, 103503 (2005)
\bibitem {ref32}
Komatsu, E., et al.: Astrophys. J. Suppl. Ser. {\bf 180}, 330 (2009)
\bibitem {ref33}
Landau, L.D., Lifshitz, E.M.: Fluid Mechanics, 2nd., Pergamon Press, Oxford, sect. 49 (1987)
\bibitem {ref34}
Maartens, R.: astro-ph/9609119 (1996)
\bibitem {ref35}
MacTavish, C.J., et al.: Astrophys. J. {\bf 647}, 799 (2006)
\bibitem {ref36}
McInnes, B.J.: High Energy Phys. {\bf 0208}, 029 (2002)
\bibitem {ref37}
Nojiri, S., Odintsov, S.D.: Gen. Rel. Grav. {\bf 38}, 1285 (2006)
\bibitem {ref38}
Nojiri, S., Odintsov, S.D.: Phys. Rev. D {\bf 72}, 023003 (2005
\bibitem {ref39}
Nojiri, S., Odintsov, S.D.: Phys Lett. B. {\bf 595}, 1 (2004)
\bibitem {ref40}
Nojiri, S., Odintsov, S.D.: Phys. Lett. B. {\bf562}, 147 (2003)
\bibitem {ref41}
Oliver, F.,  Piattella, J\'{u}lio C. Fabris., Zimdahl, W.:  JCAP. {\bf 1105}, 029 (2011)
\bibitem {ref42}
Padmanabhan, T., Chitre, S.: Phys. Lett. A. {\bf 120}, 433 (1987)
\bibitem {ref43}
Perivolaropoulos, L.: AIP Conf. Proc. {\bf 848}, 698 (2006)
\bibitem {ref44}
Perlmutter, S., et al.: Nature, {\bf 391}, 51 (1998)
\bibitem {ref45}
Riess, A.G., et al.: Astrophys. J. {\bf 607}, 665 (2004)
\bibitem {ref46}
Riess, A.G., et al.: Astron. J. {\bf 116 }, 1009 (1998)
\bibitem {ref47}
Saha, B.: Mod. Phys. Lett. A. {\bf 20}, 2127 (2005)
\bibitem {ref48}
Schmidt, B.P., et al.: Astrophys. J. {\bf 507 }, 46 (1998)
\bibitem {ref49}
Seljak, U. et al.: Phys. Rev. D. {\bf 71}, 103515 (2005)
\bibitem {ref50}
Setare, M.R., Sheykhi, A.: Int. J. Mod. Phys. D. {\bf 19}, 1205 (2010)
\bibitem {ref51}
Sheykhi, A., Setare, M.R.: Int. J. Theor. Phys. {\bf 49}, 2777 (2010)
\bibitem {ref52}
Spergel, D.N., et al.: Astrophys. J. Suppl. {\bf 148}, 175 (2003)
\bibitem {ref53}
Tegmark, M., et al., Phys. Rev. D. {\bf 69}, 103501 (2004)
\bibitem {ref54}
Tonry, J.L., et al.: Astrophys. J. {\bf 594}, 1 (2003)
\bibitem {ref55}
Zimdahl, W.: Phys. Rev. D. {\bf 53}, 5483 (1996)
\bibitem {ref56}
Zimdahl, W., Schwarz, D.J., Balakin, A.B., Pav\'{o}n, D.: Phy. Rev. D. {\bf 64}, 063501 (2001)
\end{thebibliography}
\end{document}